\documentclass[aps,twocolumn,showpacs,byrevtex,prl,reprint,nofootinbib]{revtex4-1}
\usepackage{graphicx}
\usepackage{subfigure}

\usepackage{amssymb}

\usepackage[figuresright]{rotating}

\begin{document}

\title{Light Meson Decays at BESIII
 $^*$}
 \author{Guofa XU\\
  on behalf of BESIII collaboration\\
  email: xugf@ihep.ac.cn\\
  Institute of High Energy Physics, CAS, Beijing 100049, People's Republic of China}

\affiliation{}

\begin{abstract}
The world’s largest sample of $J/\psi$ events($10^{10}$) collected at BESIII detector offers a unique opportunity to investigate $\eta$ and $\eta^{\prime}$ physics via the $J/\psi$ radiative or hadronic decays with unprecedented precision. In recent years the BESIII experiment has made significant progresses in $\eta/\eta^{\prime}$ decays. A selection of recent highlights in light meson spectroscopy at BESIII are reviewed
in this report.
\end{abstract}




\maketitle 

\section{Introduction}
After being discovered more than 50 years, the $\eta/\eta^{\prime}$ meson still attract considerable both of the theoretical and experimental attention, because of it plays a central role in our understanding of quantum chromodynamics (QCD) at low energies. As a mixture of the lowest pseudoscalar singlet and octet,
$\eta/\eta^{\prime}$ have inspired a wide variety of physics issues, e.g., $\eta- \eta^{\prime}$ mixing, the light quark masses, the fundamental discrete symmetries, as well as physics beyond the standard model (SM). Precision measurements of its decays provide important tests of effective field theories, such as chiral perturbation theory (ChPT) or the vector meson dominance (VMD) model. Moreover, it is also possible to search for new phenomena in rare or forbidden $\eta/\eta^{\prime}$ decays.

BEPCII is a double-ring multibunch $e^{+}e^{-}$ collider running in the tau-charm energy region. The BESIII detector, described in detail in Ref.\cite{bepc}, has a geometrical
acceptance of 93\% of 4$\pi$. It consists of a drift chamber (MDC), a time-of-flight (TOF) system, and an electromagnetic calorimeter (EMC), all enclosed in a superconducting solenoid with 1.0 T (0.9 T in 2012) magnetic field. The small-cell helium based MDC provides the tracking of the charged particle and ionization energy loss ($dE/dx$) measurement. The single cell position resolution is 130 $\mu m$ and the transverse momentum resolution is 0.5\% at 1 GeV/c. The TOF system for particle identification (PID) is made of plastic scintillators. It has 80 ps time resolution in the barrel, and 110 ps in the end caps. The EMC is made of 6240 CsI (Tl) crystals. The energy resolution is 2.5\% in the barrel and 5\% in the end caps for 1.0 GeV photons. Outside the solenoid, a muon chamber system made of 1272 $m^{2}$
resistive plate chambers detects muon tracks with momenta greater than 0.5 GeV\c.

The BESIII experiment has collected a total of $10^{10}$ $J/\psi$ events. Via the
$J/\psi$ radiative decay, a sample of $1.1\times 10^{7}$ $\eta$ and $5.2\times 10^{7}$ $\eta^{\prime}$ can obtained. For $\eta^{\prime}$, which is comparatively unexplored, BESIII can measure many decays for the first time and others with unrivaled precision.

\section{Observation of $\eta^{\prime}\to \pi^{+}\pi^{-}\pi^{+(0)}\pi^{-(0)}$ and $\eta^{\prime}\to 4 \pi^{0}$ }
The strong decays $\eta^{\prime}\to \pi^{+}\pi^{-}\pi^{+(0)}\pi^{-(0)}$ are not
suppressed by approximate symmetries, they are expected to be mediated by chiral anomalies, since an odd number (five) of pseudoscalar particles is involved. 

\begin{figure}[!phtb]
	\subfigure[$\eta^{\prime}\to\pi^{+}\pi^{-}\pi^{+}\pi^{-}$]{
		\label{4pi:4chrg:a}
		\begin{minipage}[h]{7.5cm}
		  \includegraphics[angle=0,width=6.5cm,height=4.5cm]{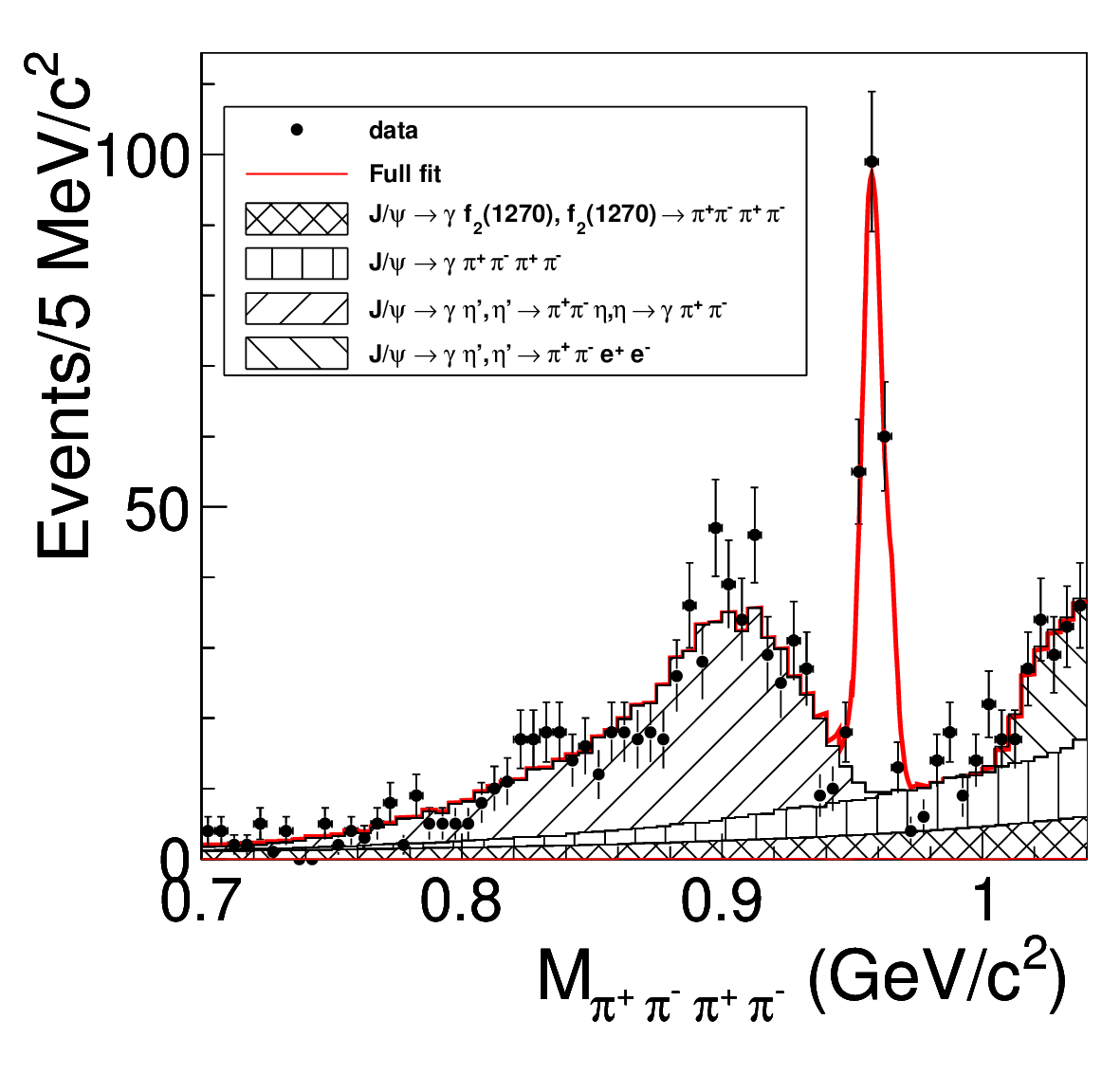}
	    \end{minipage}}%
	\hfill
	\subfigure[$\eta^{\prime}\to\pi^{+}\pi^{-}\pi^{0}\pi^{0}$]{
		\label{4pi:2c2n:b} 
		\begin{minipage}[h]{7.5cm}
		  \includegraphics[angle=0,width=6.5cm,height=4.5cm]{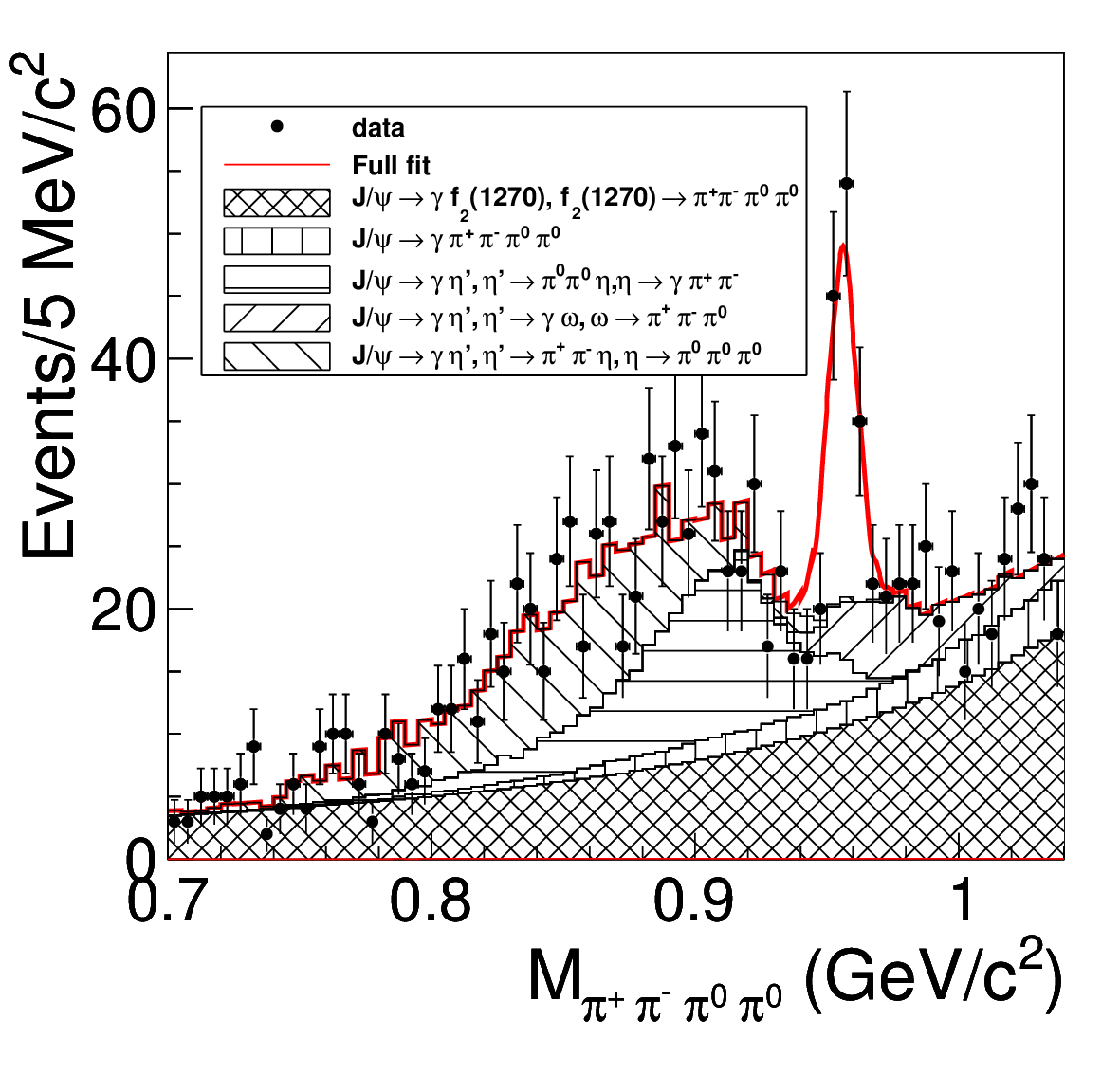}
	    \end{minipage}}%
	\hfill
	\subfigure[$\eta^{\prime}\to 4\pi^{0}$]{
	\label{4pi:4neu:c} 
	\begin{minipage}[h]{7.5cm}
		\includegraphics[angle=0,width=6.5cm,height=4.5cm]{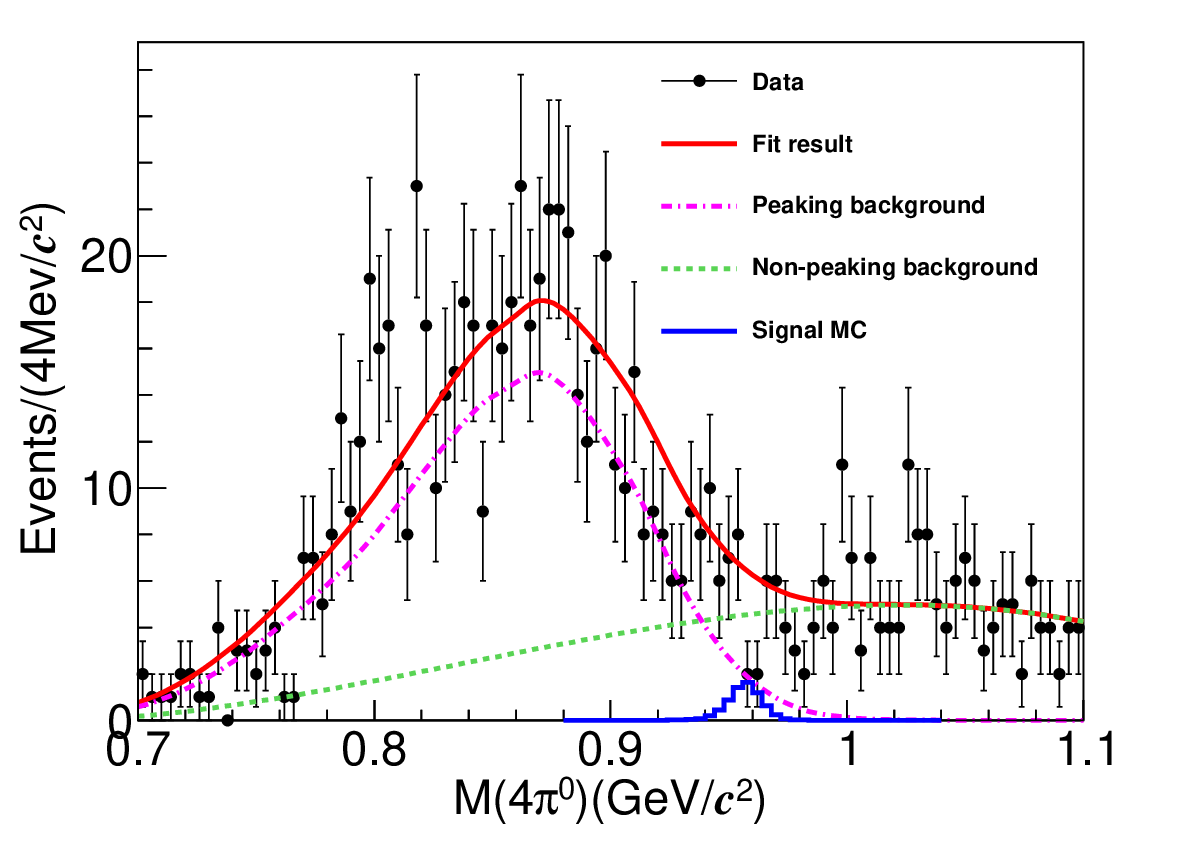}
    \end{minipage}}   
	\caption{The invariant mass spectrum of (a) $\pi^{+}\pi^{-}\pi^{+}\pi^{-}$, (b) $\pi^{+}\pi^{-}\pi^{0}\pi^{0}$, and (c) $4\pi^{0}$ for data (black dot with error bars), together with the total fit result (red curve).} 
	\label{4pi}
\end{figure}

The projections of the fit to $M_{\pi^{+}\pi^{-}\pi^{+(0)}\pi^{-(0)}}$
in the $\eta^{\prime}$ mass region are shown in Figs.~\ref{4pi:4chrg:a} and ~\ref{4pi:2c2n:b}, where the shape of the sum of signal and background shapes are in good agreement with data. We obtain $199\pm 16$ of $\pi^{+}\pi^{-}\pi^{+}\pi^{-}$ events with a statistical significance of $18\sigma$ and $84\pm 16$ of $\pi^{+}\pi^{-}\pi^{0}\pi^{0}$ events with a statistical significance of $5\sigma$~\cite{4pi:c}. Using the world average value of $Br(J/\psi\to\gamma\eta{\prime})$~\cite{PDG}, the branching fraction of $\eta^{\prime}\to\pi^{+}\pi^{-}\pi^{+(0)}\pi^{-(0)}$ are determined to be $Br(\eta^{\prime}\to\pi^{+}\pi^{-}\pi^{+}\pi^{-}=[8.53\pm 0.69(stat.)\pm 0.64(syst.)]\times 10^{-5})$ and $Br(\eta^{\prime}\to\pi^{+}\pi^{-}\pi^{0}\pi^{0}=[1.82\pm 0.35(stat.)\pm 0.18(syst.)]\times 10^{-4})$, which are consistent with the theoretical predictions based on a combination of ChPT and VMD models, but not with the broken-SU$_{6}\times O_{3}$ quark model~\cite{4pi:chpt}. 

 In theory, $\eta^{\prime}\to 4 \pi^{0}$ is a highly suppressed decay because of the S-wave CP-violation. In the light of an effective chiral Lagrangian approach, the S-wave CP-violation in $\eta^{\prime}\to 4\pi^{0}$ is induced by the so-called $\theta$-term, which is an additional term in the QCD Lagrangian to account for the
 solution of the strong-CP problem. It was found that the S-wave CP-violation effect that contributed to this decay is at a level of $10^{-23}$~\cite{4pi0}. While higher-order contributions, involving a D-wave pion loop or the production of two $f_{2}$ tensor mesons, provide a CP-conserving route through which the decay can occur. By ignoring the tiny
 contribution from the latter process, calculations based on ChPT and VMD model predict the branching fraction caused
 by D-wave CP-conserving to be at the level of $10^{-8}$~\cite{4pi0:1}. However, the theoretical prediction is not strictly based on the effective field theory due to the lack of knowledge at such a high order in the chiral expansion and the use of a model to make an estimation. One does not know the reliability of that model a priori. Therefore, a search for the decay  $\eta^{\prime}\to 4 \pi^{0}$ is
 useful to check the reliability of it.

Figure~\ref{4pi:4neu:c} shows the $4\pi^{0}$ invariant mass spectrum in data, together with the total fit result and the contributions from non-peaking background and the peaking background $J/\psi\to\gamma\eta^{\prime}, \eta^{\prime}\to\pi^{0}\pi^{0}\eta, \eta\to 3\pi^{0}$. Also shown is the expected shape of the signal contribution, with arbitrary normalization. After our selection, there are no significant $\eta^{\prime}$ signal is evident. With a Bayesian approach, the upper limit on the branching fraction is determined to be $Br(\eta^{\prime}\to 4\pi^{0}<4.94\times 10^{-5})$ at $90\%$ C.L.~\cite{4pi:n}. This corresponds to an improvement of a factor 6 compared to the previous best value from the GAMS-4$\pi$ experiment~\cite{4pi0:2}. However, the current limit is still far to reach the theoretical predication with a level of $10^{-8}$.

\section{Observation of $\eta^{\prime}\to\rho^{+}\pi^{-}+c.c.$}
The decays $\eta^{\prime}\to\pi\pi\pi$ are isospin-violating processes. Because the electromagnetic contribution is strongly suppressed~\cite{rhopi:1}, they are induced dominantly by the strong interaction via the explicit breaking of chiral symmetry by
the $d-u$ quark mass difference. 

A Dalitz plot analysis based on the formalism of the isobar model~\cite{rhopi:2} is performed. The resonant $\pi-\pi$ S-wave ($L=0$ for $\sigma$) and P-wave ($L=1$ for $\rho^{\pm}$) amplitudes are described as in Ref.~\cite{rhopi:3}.

\begin{figure}[!htbp]
	\centering
	\includegraphics[width=0.25\textwidth]{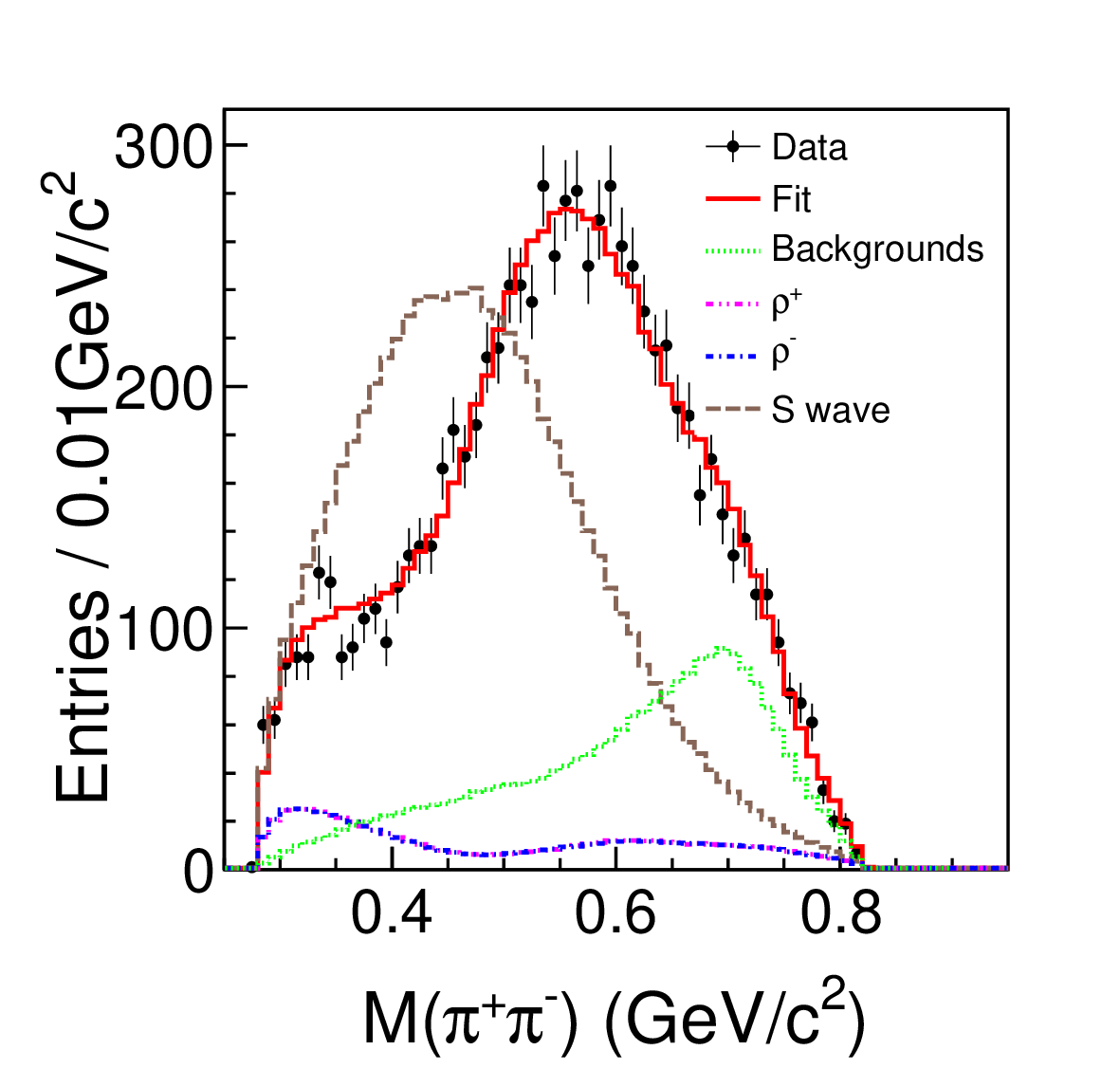}\put(-85,85){\bf (a)}
	\includegraphics[width=0.25\textwidth]{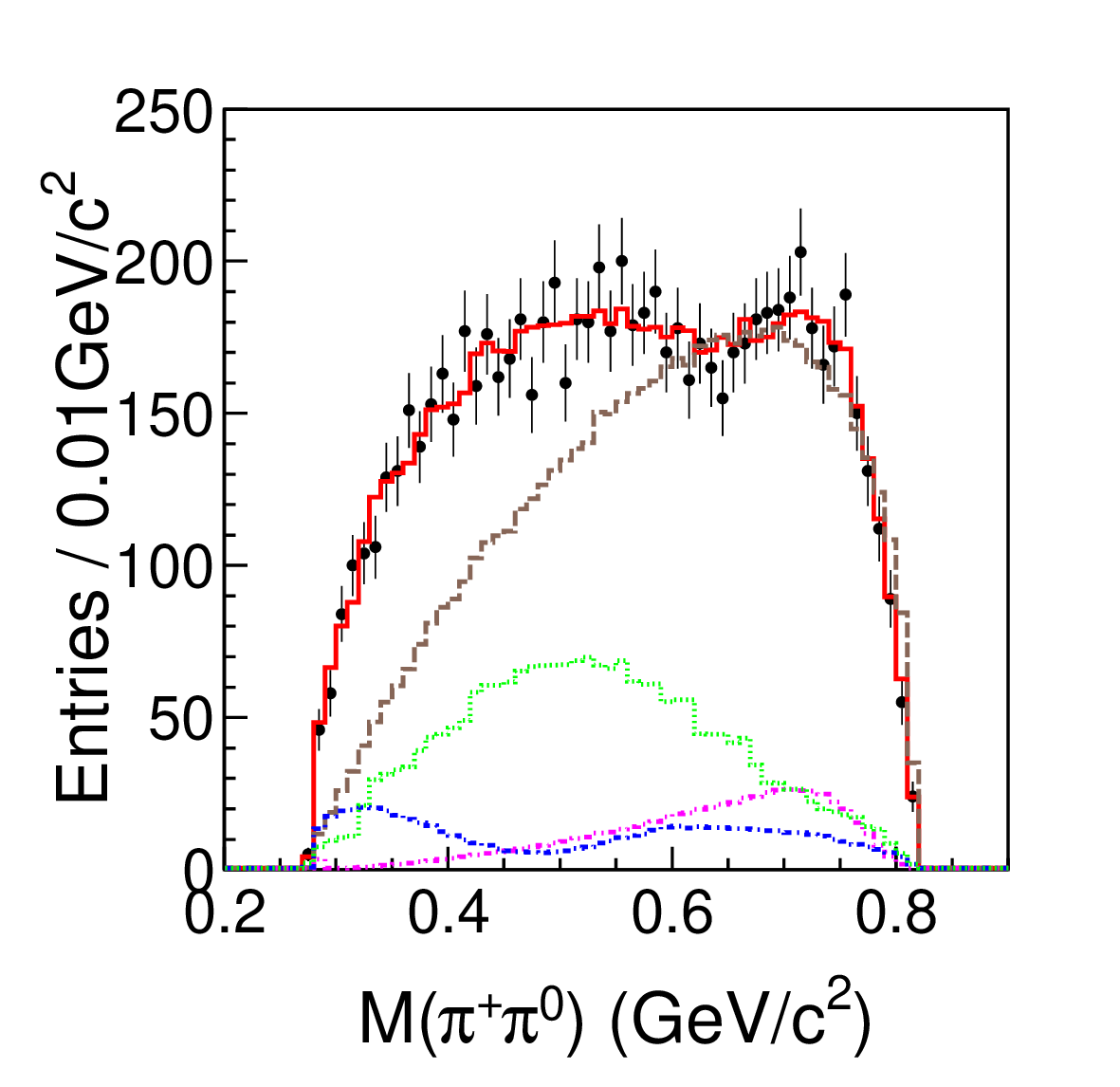}\put(-85,85){\bf (b)}
	
	\includegraphics[width=0.25\textwidth]{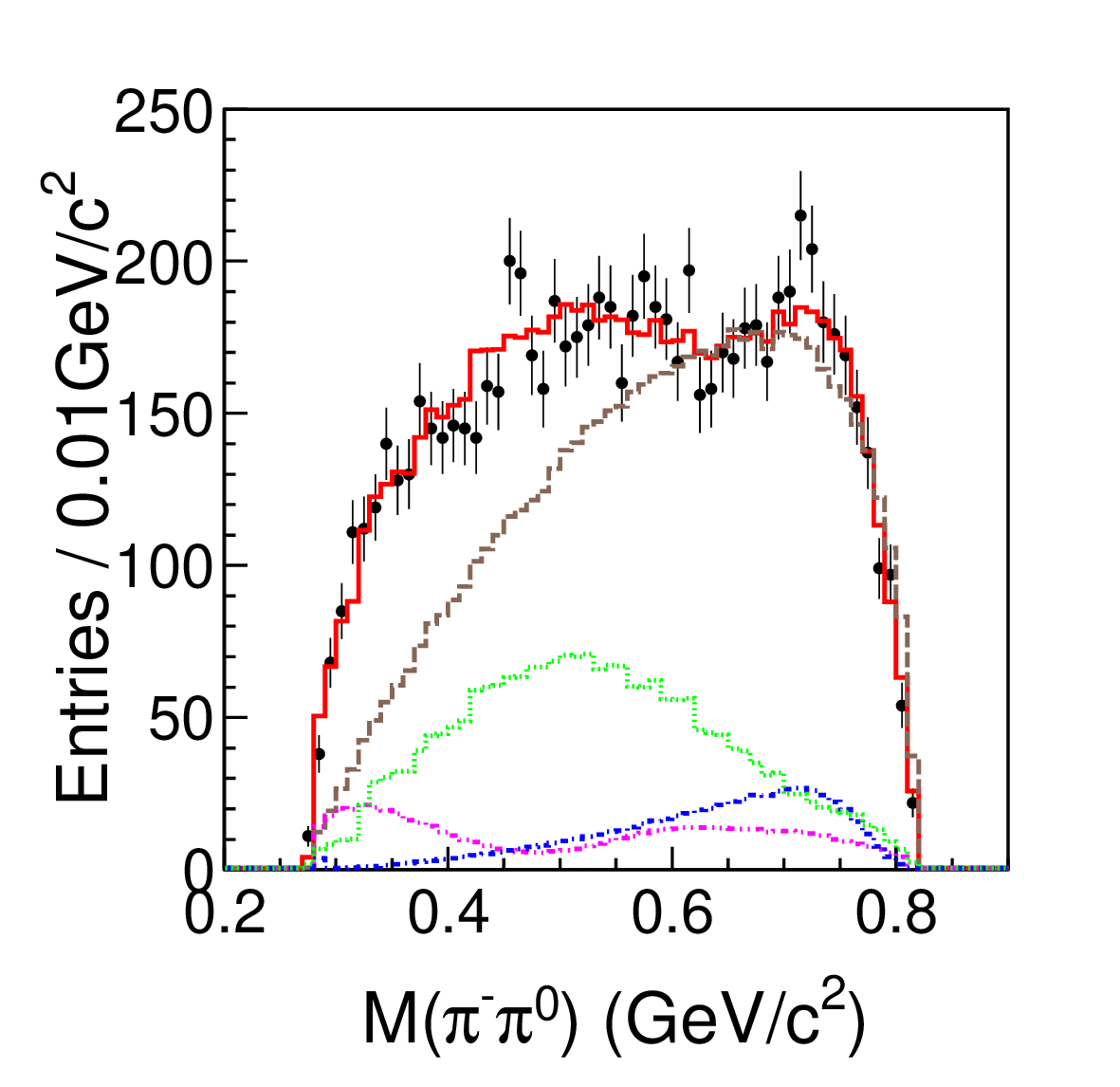}\put(-85,85){\bf (c)}
	\includegraphics[width=0.25\textwidth]{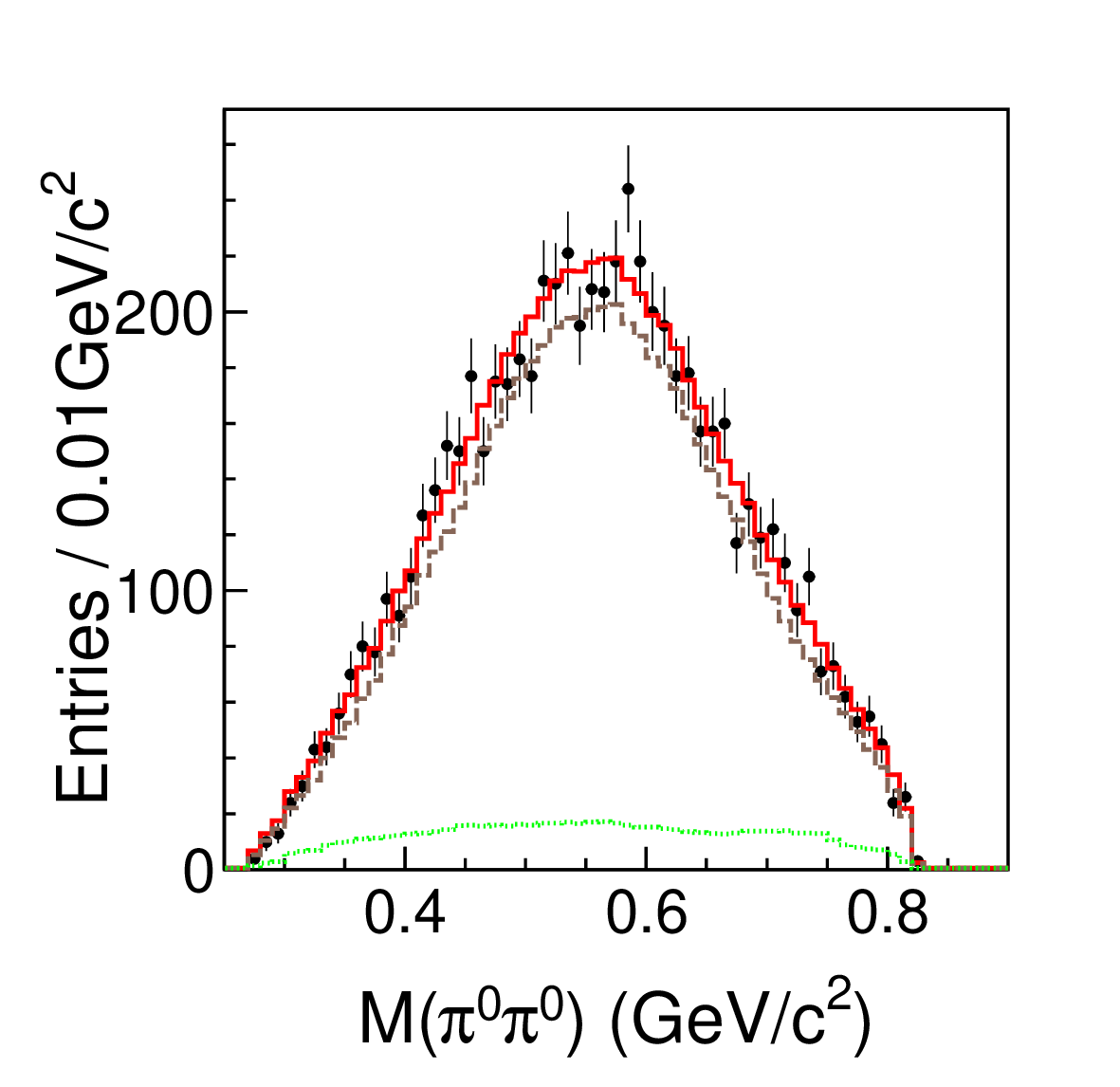}      \put(-85,85){\bf (d)}
	\caption{\label{fig:rhopi} Comparison of the invariant mass distributions of (a) $\pi^{+}\pi^{-}$,
		(b) $\pi^{+}\pi^{0}$, (c) $\pi^{-}\pi^{0}$, and (d) $\pi^{0}\pi^{0}$ between data (dots with error bars)
		and the fit result projections (solid histograms).
		The dotted, dashed, dash-dotted, and dash-dot-dotted histograms show
		the contributions from background, $S$ wave, $\rho^{-}$, and $\rho^{+}$, respectively.
	}
\end{figure}

Projections of the data and fit results are displayed in Fig.~\ref{fig:rhopi}. The data are well described by three
components: $P$ wave ($\rho^\pm\pi^\mp$), resonant $S$ wave ($\sigma\pi^{0}$), and phase-space $S$ wave ($\pi\pi\pi$).  The
interference between $\sigma$ and the non-resonant term is large and strongly depends on the parametrization of $\sigma$. Therefore we are unable to determine the individual contributions and consider only the sum of the $S$-wave amplitudes in this analysis.

Using a combined amplitude analysis of $\eta^{\prime}\to\pi^{+}\pi^{-}\pi^{0}$ and $\eta^{\prime}\to\pi^{0}\pi^{0}\pi^{0}$ decays, the $P$-wave contribution from $\rho^\pm$ is observed for the first time
with high statistical significance.
The pole position of $\rho^\pm$, $775.49$ (fixed)$-i(68.5\pm0.2)$ MeV, is consistent with previous measurements, and the branching fraction ${\mathcal Br}(\eta^{\prime}\to\rho^{\pm}\pi^{\mp})$
is determined to be $(7.44\pm0.60\pm1.26\pm1.84)\times 10^{-4}$~\cite{rhopi}.

\section{Observation of the Dalitz Decay $\eta^{\prime}\to\gamma e^{+}e^{-}$}

Electromagnetic (EM) Dalitz decays of light pseudoscalar mesons, $P\to \gamma l^{+}l^{-}$ ($P=\pi^{0}$, $\eta$, $\eta^{\prime}$; $l=e, \mu$), play an
important role in revealing the structure of hadrons and the interaction mechanism between photons and hadrons~\cite{gee:1}.  If one assumes point-like particles, the decay rates can be exactly calculated
by Quantum Electrodynamics (QED)~\cite{gee:2}. Modifications to the QED decay rate due to the inner structure of
the mesons are encoded in the transition form factor (TFF) $F(q^2)$, where $q$ is the momentum transferred to the lepton pair, and $q^2$ is the square of the invariant mass of the lepton pair. A recent summary and discussion of this subject
can be found in Ref.~\cite{gee:3}.

We report the first observation of the $\eta^{\prime}\to \gamma e^{+}e^{-}$ decay
and the extraction of the TFF. The source of the $\eta^{\prime}$ mesons
are from radiative $J/\psi\to\gamma\eta^{\prime}$ decays collected by the BESIII at the BEPCII $e^{+}e^{-}$ collider. The $\eta^{\prime}\to\gamma\gamma$ decay events
in the same data sample are used for normalization.

\begin{figure}[hbtp]
	\centering
	\includegraphics[width=0.8\linewidth]{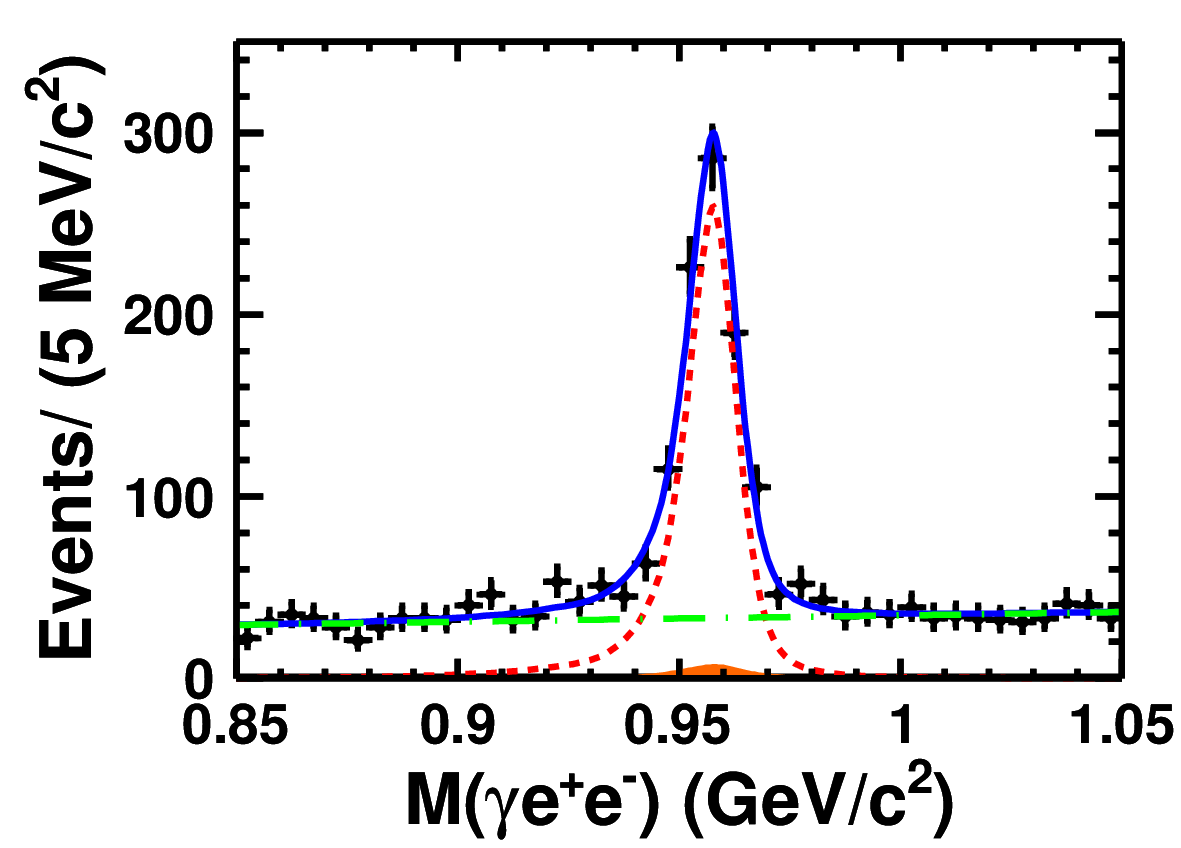}
	\caption{Invariant $\gamma e^{+}e^{-}$ mass distribution for the selected signal events. The (black) crosses
	are the data, the (red) dashed line represents the signal, the (green) dot-dashed curve shows the non-peaking background shapes, the (orange) shaded component is the shape of the
	$J/\psi \to \gamma \eta^{\prime}, \eta^{\prime}\to\gamma \gamma$ peaking
	background events. The total fit result is shown as the (blue) solid line.}
	\label{gee:etapfit}
\end{figure}

\begin{figure}[hbtp]
	\centering
	\includegraphics[width=0.8\linewidth]{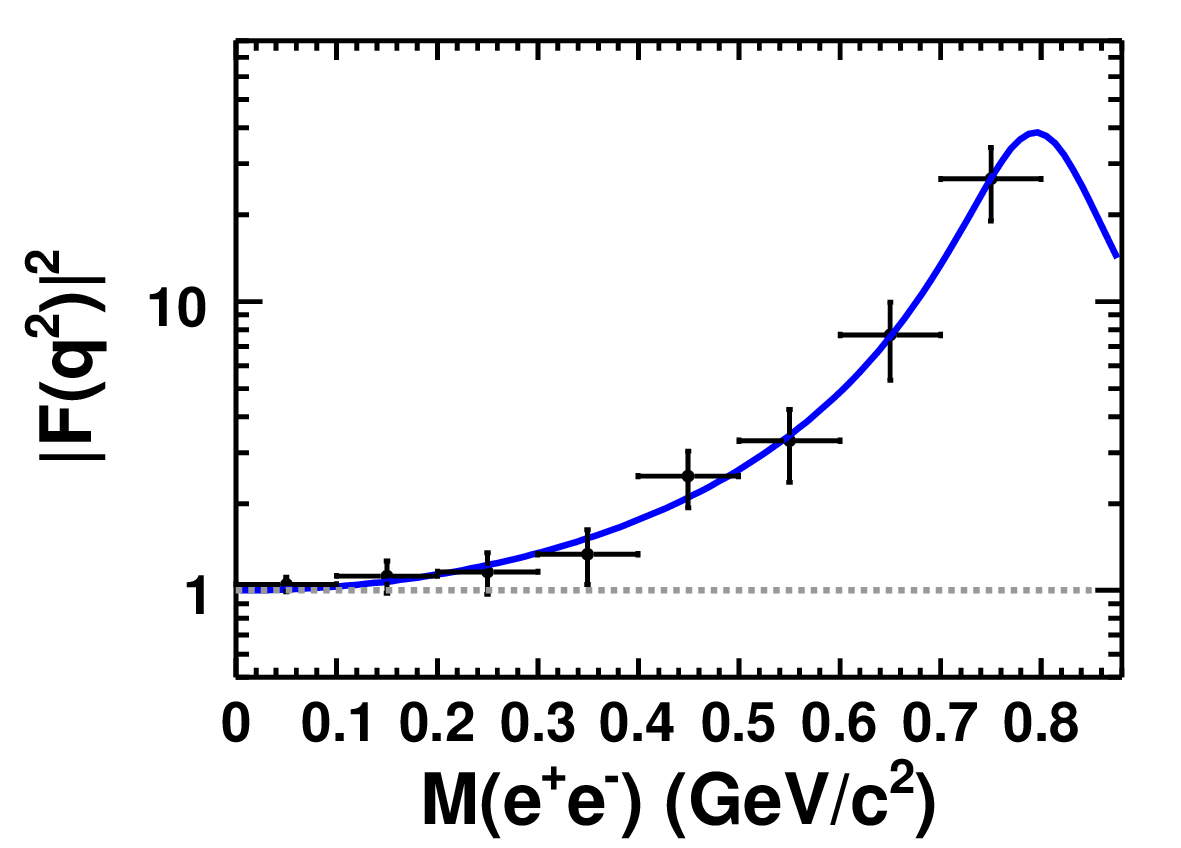}
	\caption{Fit to the single pole form factor  $|F|^2$. The (black) crosses
	are data, where the statistical and systematic uncertainties are combined, the (blue) solid curve shows the fit results. The (gray) dotted line shows to the point-like case (\emph{i.e.} with $|F|^2=1$) for comparison.}
	\label{gee:TFFfit}
\end{figure}

The combination of $\gamma e^{+}e^{-}$ with invariant mass closest to $m_{\eta^{\prime}}$ is taken to reconstruct the $\eta^{\prime}$. The resulting $M(\gamma e^{+}e^{-})$ distribution after the selection criteria is shown in Fig.~\ref{gee:etapfit} and exhibits a clear peak at the $\eta^{\prime}$ mass. An unbinned extended maximum likelihood fit is performed to determine the signal yield. Using the  $\eta^{\prime}\to \gamma \gamma$ branching fraction value listed in PDG~\cite{PDG}, we obtain the first measurement of the $\eta^{\prime} \to \gamma e^{+}e^{-}$ branching fraction of
${Br}(\eta^{\prime} \to \gamma e^{+}e^{-})= (4.69\pm0.20(stat.)\pm0.23(sys.))\times10^{-4}.$~\cite{gee:0}

The results of a least-squares fit
with the single pole model is shown in Fig.~\ref{gee:TFFfit}, the parameters of the
form factors are determined to be
$\Lambda_{\eta^{\prime}} = (0.79\pm0.05)$~GeV, $\gamma_{\eta^{\prime}} = (0.13\pm0.06)$~GeV. From the fitted value of the parameter $\Lambda_{\eta^{\prime}}$, the slope of the form factor is obtained
to be $(1.60\pm0.19)$~GeV$^{-2}$~\cite{gee:0}, in agreement with the result
$b_{\eta^{\prime}} = (1.7\pm0.4)$ GeV$^{-2}$ obtained in the process of
$\eta^{\prime} \to \gamma \mu^{+}\mu^{-}$~\cite{gee:1}.

\section{Observation of the double Dalitz decay $\eta^{\prime}\to e^+e^-e^+e^-$}
The double Dalitz decays $P\to \ell^+ \ell^- \ell'^+\ell'^-$, where $P$ is a pseudoscalar meson ($P=\pi^0, \eta$, or $\eta'$) while $\ell$ and $\ell'$ are leptons ($\ell, \ell'=e,\mu$), are expected to proceed through an intermediate state of two virtual photons. These processes are of great interest for understanding the
pseudoscalar TFF and the interactions between pseudoscalars and virtual photons~\cite{4e:1}. These TFFs are necessary inputs to calculate the pseduoscalar-meson-pole contributions to the hadronic light-by-light scattering, which causes the second largest uncertainty in the Standard Model determination of the muon anomalous magnetic moment~\cite{4e:2}.
Particularly, the double Dalitz decays of pseudoscalar mesons help to determine the TFFs in the small timelike momentum region, i.e. $m_{ll}^2\le q^2\le m_P^2$, with $m_{ll}$ the invariant mass of the dilepton and $m_P$ the mass of the pseudoscalar meson, and thus are suitable
to determine the slope of the TFFs at $q^2=0$~\cite{4e:2}.

 \begin{figure}[!http] 
	\centering
	\includegraphics[width=0.5\textwidth]{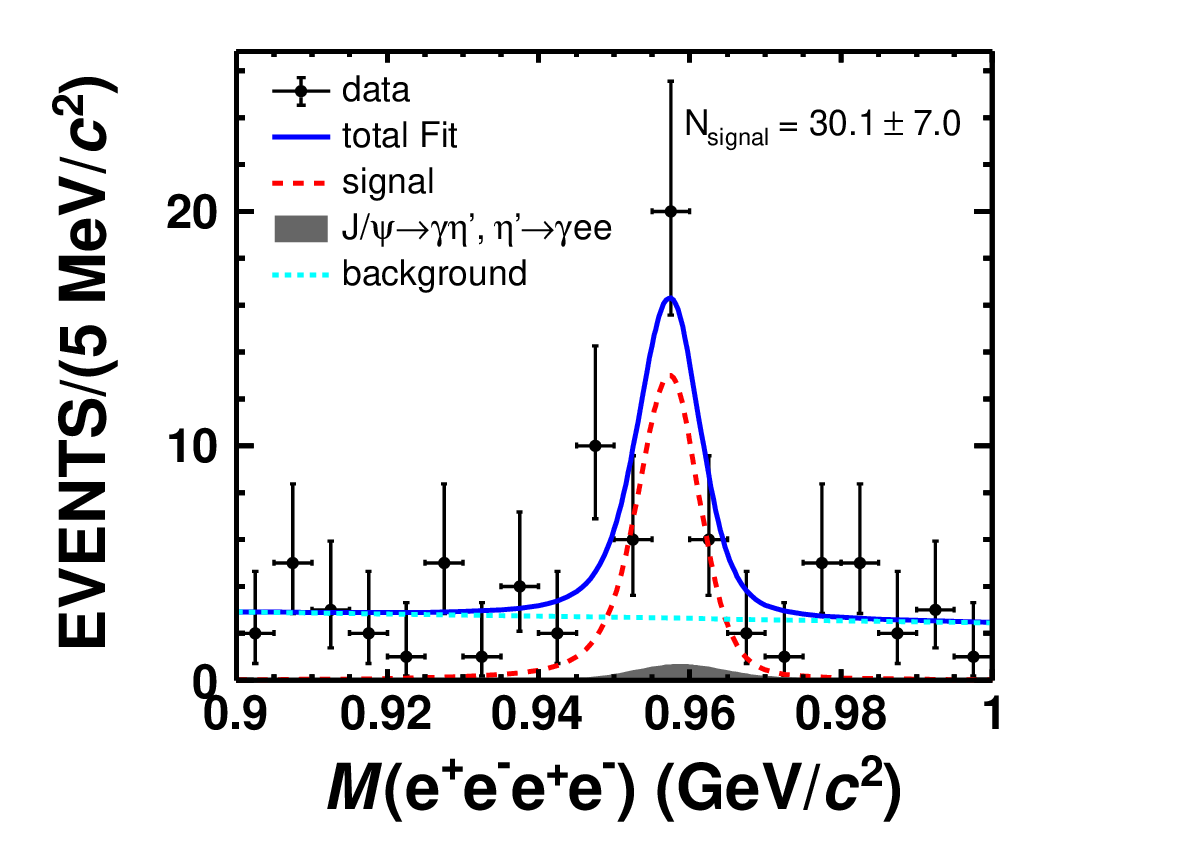} 
	\caption{The $M(e^+
		e^- e^+ e^-)$ distribution of data and the fitting results. The
		dots with error bars are data, the red dashed line is the signal
		shape, and the solid blue line is the total fit result. The gray area
		represents the peaking background from $J/\psi\to \gamma\eta',
		\eta'\to\gamma e^+e^-$, and the cyan dotted line is a linear
		function. }
	\label{m4e:fit}
\end{figure}

The resulting $M(e^+ e^- e^+ e^-)$
distribution is shown in Fig.~\ref{m4e:fit}, where a clear $\eta'$ signal is visible. An unbinned extended maximum likelihood fit is performed
to determine the $\eta'$ signal yield. The fit results shows the $\eta'$ signal with a significance of 5.7$\sigma$, and the calculated branching fraction is $Br(\eta'
\to e^+e^-e^+e^-)= (4.5\pm1.0(stat.)\pm0.5(sys.)) \times 10^{-6}$~\cite{4e:0}, it is consistent with the theoretical predictions within the
uncertainties and provides new information for the studies about $\eta'$ TFF and the interactions between $\eta'$ and virtual
photons~\cite{4e:1}.

\section{Study of $\eta'\to\gamma\pi^{+}\pi^{-}$ Decay Dynamics}
The radiative decay $\eta' \to \gamma\pi^+ \pi^-$ is the second most
probable decay mode of the $\eta'$ meson with a branching fraction of
$(28.9\pm0.5)$\%~\cite{PDG} and is frequently used for tagging $\eta'$ candidates. In the VMD model, this process is dominated by the decay $\eta' \to\gamma\rho^0$. In the past, the dipion mass distribution was studied by several experiments, and a peak shift of about $+20~{\rm MeV}$/$c^2$
for the $\rho^0$ meson with respect to the expected position was
observed. Dedicated studies, concluded that a lone $\rho^0$ contribution in the
dipion mass spectrum did not describe the experimental data~\cite{gpipi:1}. This discrepancy could be attributed to a higher
term of the Wess-Zumino-Witten anomaly, known as the box anomaly, in the ChPT Lagrangian~\cite{gpipi:2}. Both the model-dependent and model-independent approaches are carried out to investigate the decay dynamics. 

\begin{figure}[htbp]
	\centering
	\includegraphics[width=7.5cm, height=6.5cm]{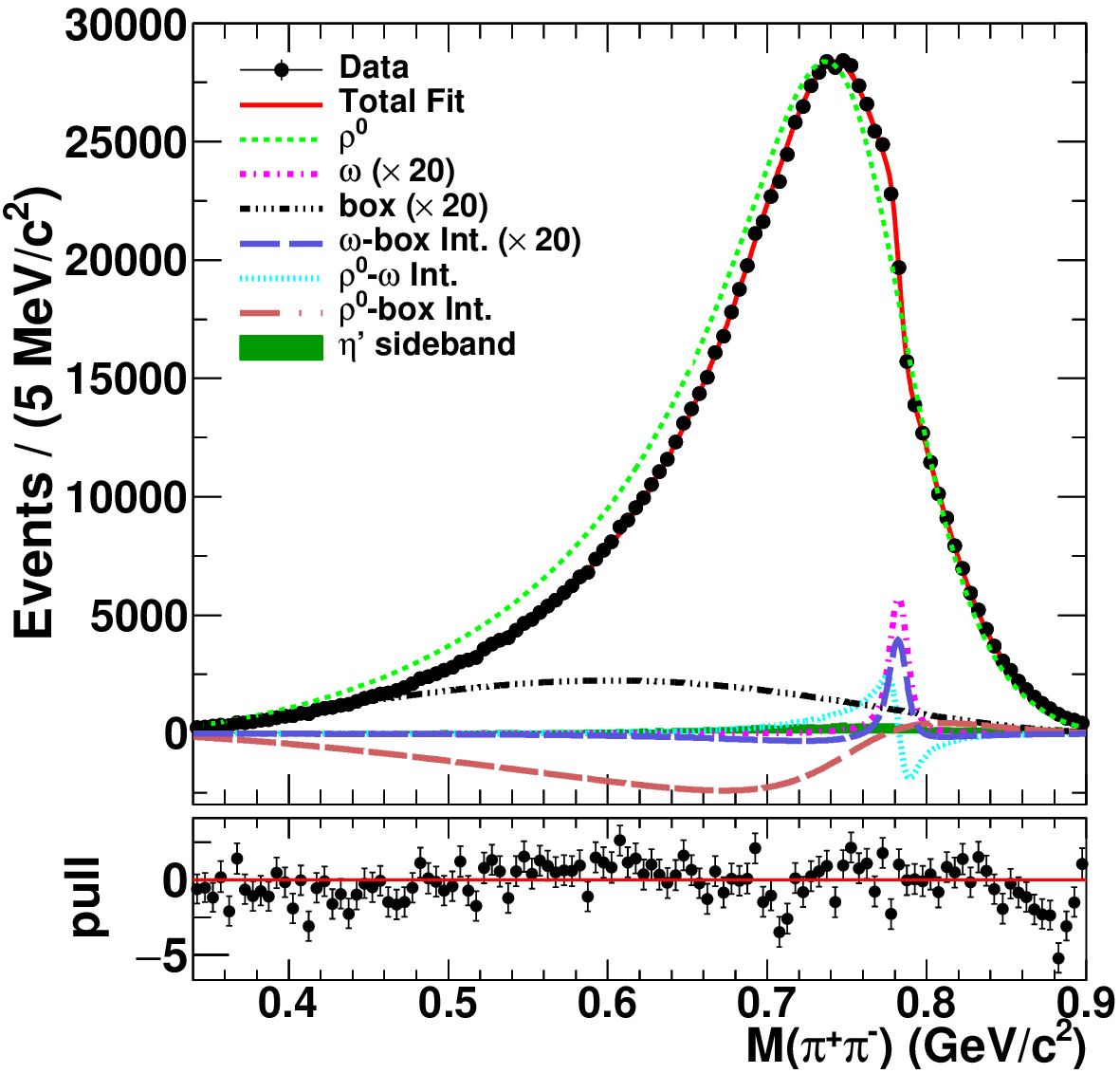}
	\put(-40,160){\bf \large~(a)}
	
	\includegraphics[width=7.5cm, height=6.5cm]{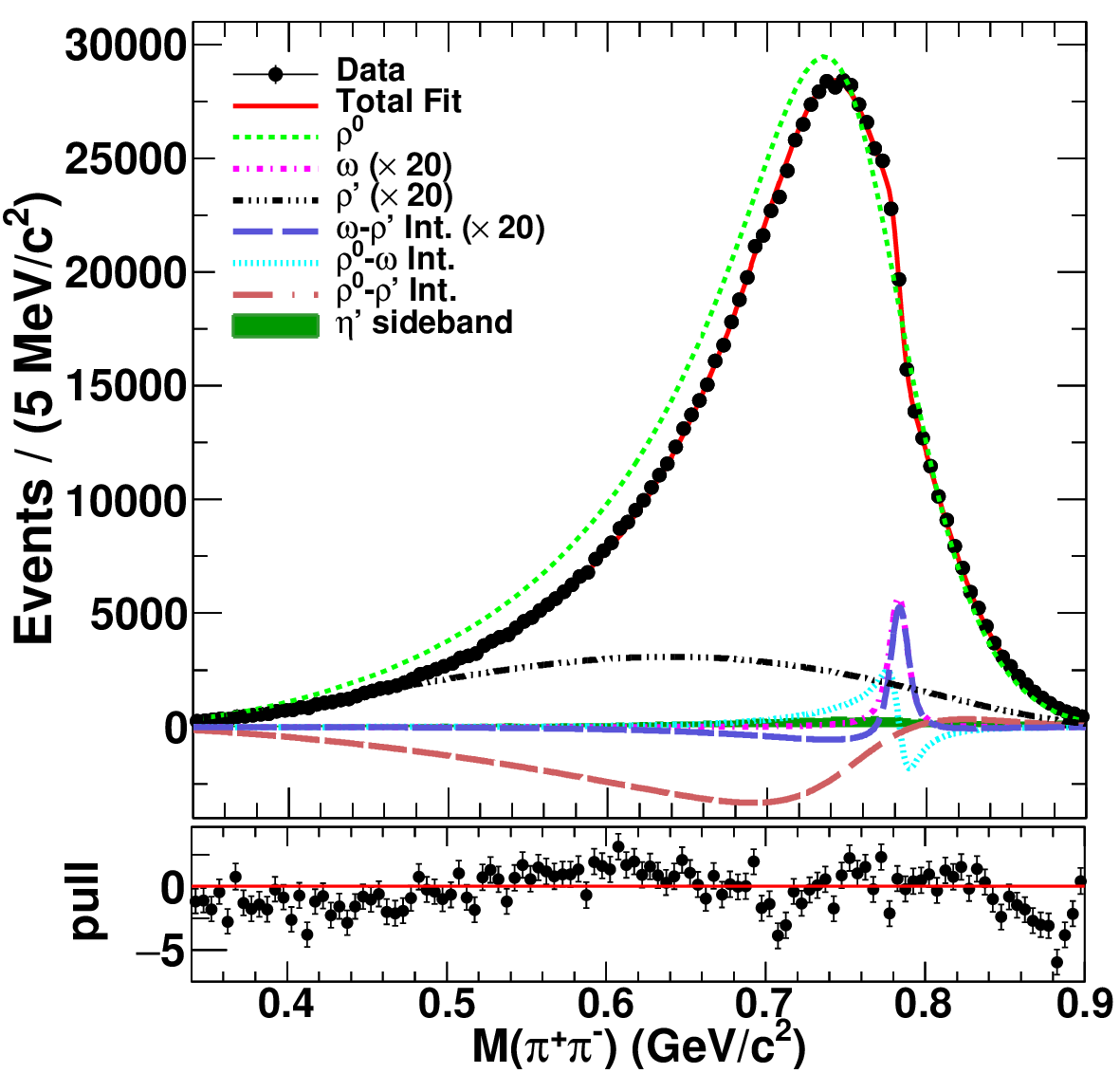}
	\put(-40,160){\bf \large~(b)}
	\caption{Model-dependent fit results in case (a)
		$\rho^0$-$\omega$-box anomaly and (b)
		$\rho^0$-$\omega$-$\rho^{\prime}$. Dots with error bars represent
		data, the green shaded histograms are the background from $\eta'$
		sideband events, the red solid curves are the total fit results,
		and others represent the separate contributions as indicated. To be
		visible, the small contributions of $\omega$, the box anomaly
		($\rho^{\prime}$) and the interference between $\omega$ and the box anomaly ($\rho^{\prime}$) are scaled by a factor of 20.}
	\label{gpipi:dependent}
\end{figure}

In the model-dependent study, binned maximum likelihood fits are performed to the $M({\pi^{+}\pi^{-}})$ distribution between 0.34 and 0.90 GeV$/c^2$ with different scenarios. Finally, we find the ``$\rho^0$-$\omega$-box anomaly" model gives the best goodness of fit $\chi^2/ndf$ = 207/107 (Fig.~\ref{gpipi:dependent}(a))~\cite{gpipi:0}, An
alternative fit by replacing the box anomaly with the $\rho^{\prime}$
component gives considerably worse agreement with $\chi^2/ndf$ = 303/106 (Fig.~\ref{gpipi:dependent}(b)).  

A model independent approach is also implemented to investigate the decay dynamics. The model independent approach  provides a satisfactory parametrization of the dipion invariant mass spectrum,
and yields the parameters of the process-specific part $P(s)$ to be
$\kappa=0.992\pm0.039\pm0.067\pm0.163$ GeV$^{-2}$,
$\lambda=-0.523\pm0.039\pm0.066\pm0.181$ GeV$^{-4}$, and
$\xi=0.199\pm0.006\pm0.011\pm0.007$, where the first uncertainties are statistical, the second are systematic, and the third are
theoretical. In contrast to the conclusion in Ref.~\cite{gpipi:3} based on the limited statistics from the Crystal Barrel experiment~\cite{gpipi:4}, our result indicates that the quadratic term and the $\omega$ contribution in $P(s)$, corresponding to statistical significances of $13\sigma$ and $34\sigma$, respectively, are necessary. 

\begin{figure}[htbp]
	\includegraphics[width=7.5cm, height=6.5cm]{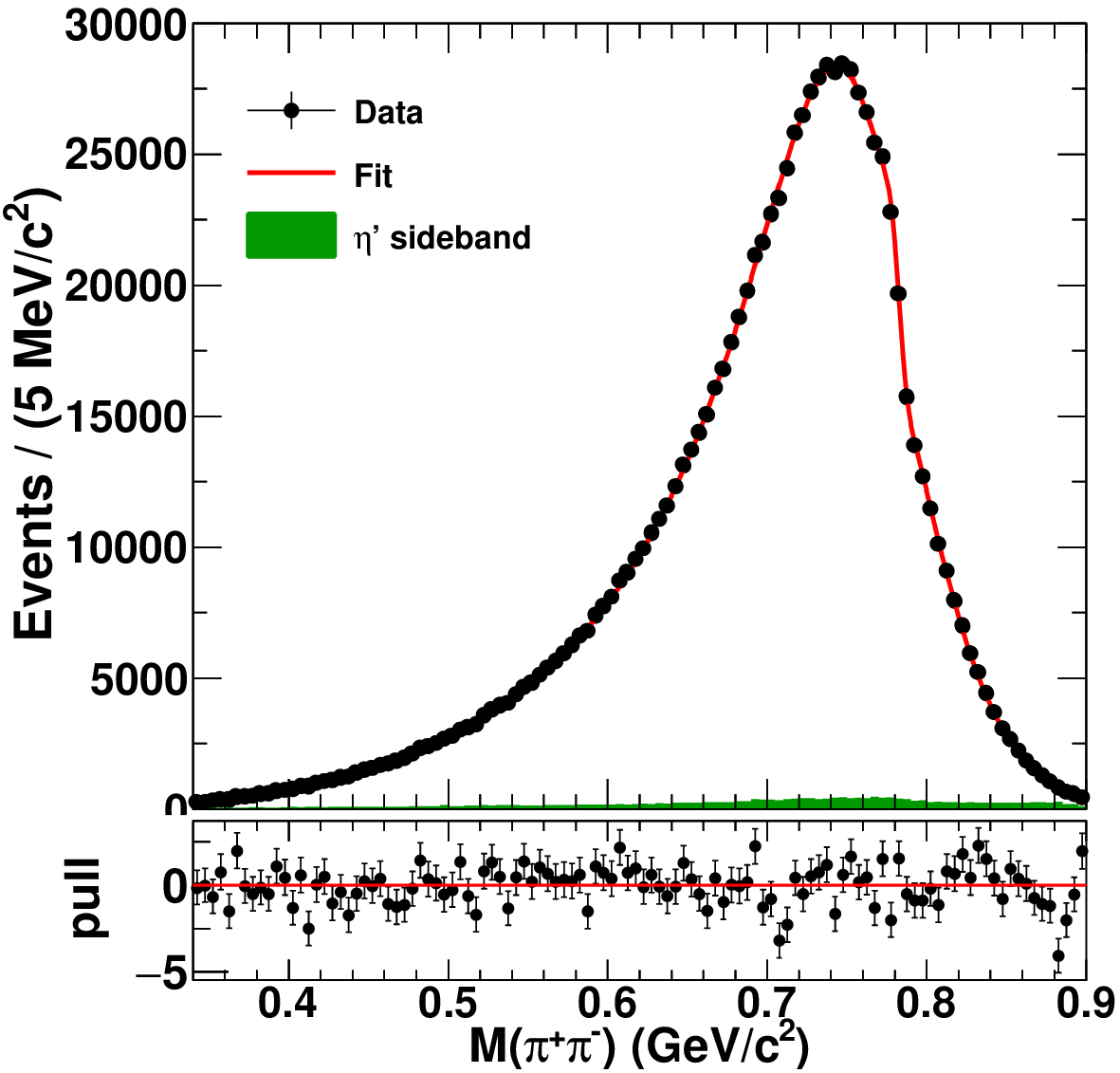}
	\caption{The results of the model-independent fit with
		$\omega$ interference. Dots with error bars represent data, the
		(green) shaded histogram is the background contribution from $\eta^{\prime}$
		side-band events, and the (red) solid curve is the fit result.}
	\label{gpipi:independent}
\end{figure}

\section{Evidence for the Cusp Effect in $\eta'$ Decays into $\eta\pi^0\pi^0$}
Experimental studies of light meson decays are important guides to our understanding of how QCD works in the non-perturbative regime. In $\pi\pi$ interaction, one of the prominent features is the loop contribution to the $\pi\pi$ scattering: the $S$-wave charge-exchange rescattering $\pi^+\pi^-\to\pi^0\pi^0$ causes a prominent cusp at the center of mass energy
corresponding to the summed mass of two charged pions. The cusp effect can shed
light on the fundamental properties of QCD at low energies, by determining the
strength of the $S$-wave $\pi\pi$ interaction.
\begin{figure}
	\centering
	\includegraphics[width=7.5cm,     
	          height=6.5cm]{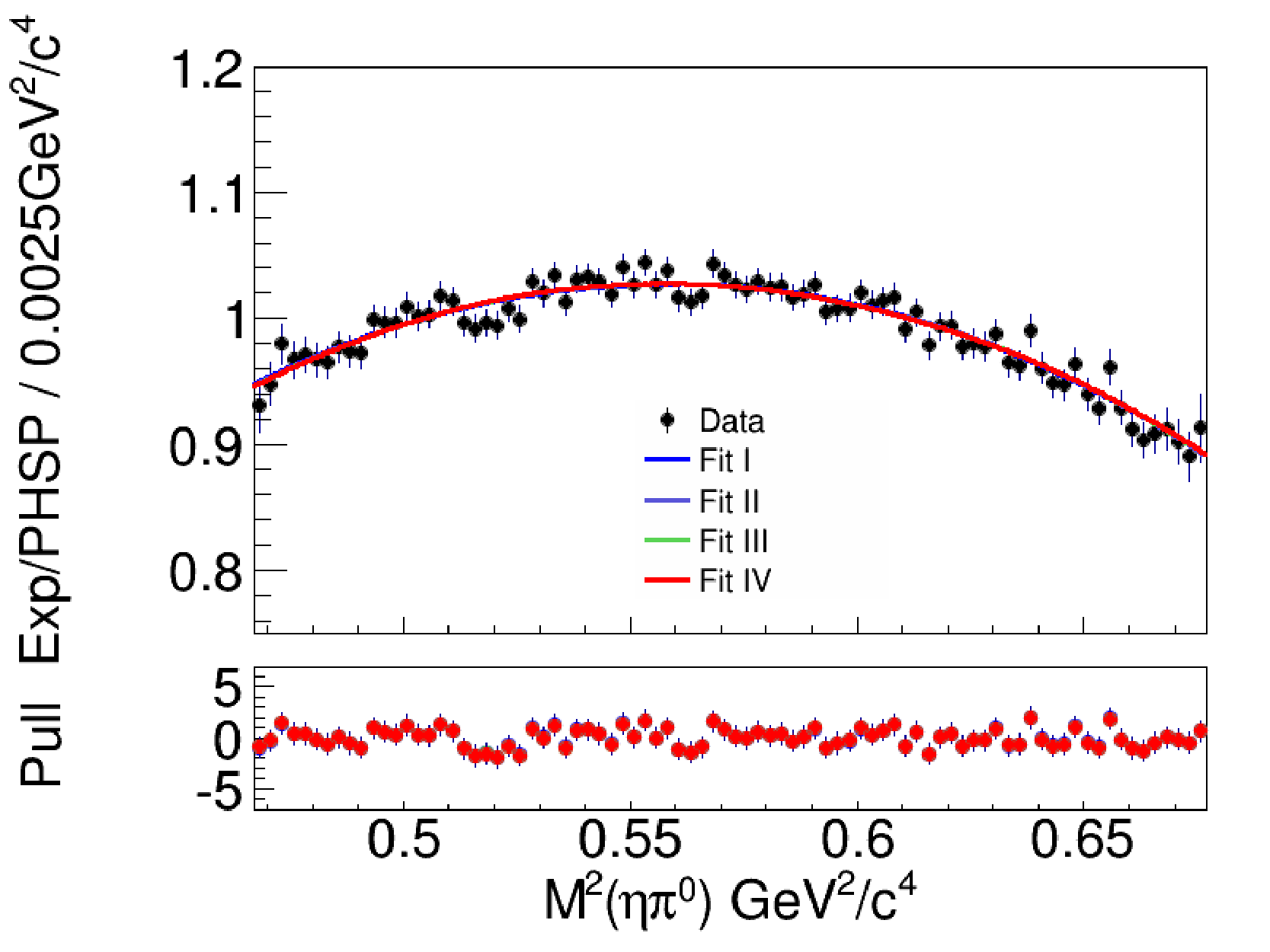}
      \put(-40,160){\bf \large~(a)}
      
    \includegraphics[width=7.5cm, 
              height=6.5cm]{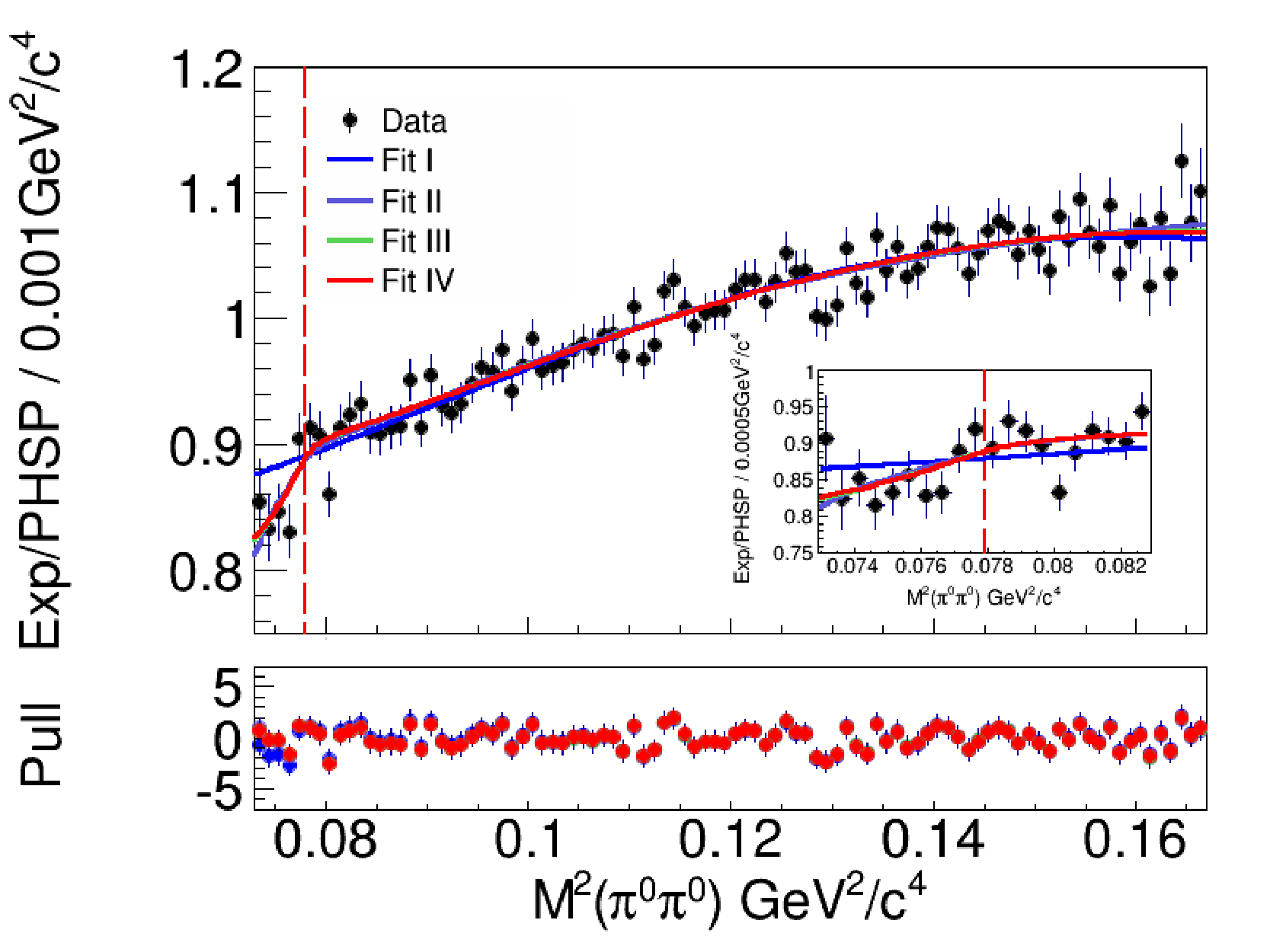}
      \put(-40,160){\bf \large~(b)}
	\caption{The fit result projections divided by phase space of different models to variable 
		(a)$M^{2}(\eta\pi^0)$ and (b)$M^{2}(\pi^0\pi^0)$. 
	The black dots with error bars are
	from data. The solid lines are fit results from the corresponding models. The red dashed line indicates the charged pion mass threshold. The cusp region is also shown in the inset.}
	\label{fig:eta2pi0}
\end{figure}

Using an unbinned maximum likelihood method, we fit the Dalitz plot of
$M^2(\pi^0\pi^0)$ versus $M^2(\eta\pi^0)$ within the framework of NREFT. The resolution effect and detection efficiency are studied by MC simulation and taken into account in the fit. We perform alternative fits within the framework of NREFT to evaluate this effect (Fig.~\ref*{fig:eta2pi0}). The fit with tree level amplitude (Fit I) shows a discrepancy below the charged pion mass threshold, which implies the existence of the cusp effect. To describe the data in this region, the contributions at one- and two-loop level (Fit II $\sim$ Fit IV) are introduced in the decay amplitude. We perform alternative analyses by taking into account the cusp effect. For each
case, the amplitude provides a good description of the structure around the
charged pion mass threshold and the statistical significance is found to be
around $3.5\sigma$~\cite{eta2pi0:0}. The scattering length combination $a_0-a_2$ is measured to
be $0.226\pm0.060\pm 0.013$, which is in good agreement with the theoretical
value of $0.2644\pm0.0051$~\cite{eta2pi0:1} within the uncertainties. The observation of the evidence of the cusp effect in $\eta'\to\eta\pi^0\pi^0$ decay demonstrates the excellent potential to investigate the underlying dynamics of light mesons at the BESIII experiment.

\section{First Measurement of Absolute Branching Fractions of $\eta/\eta'$ Decays}
As two members of the ground-state nonet of pseudo-scalar mesons, the $\eta$ and $\eta'$ mesons play an important part in understanding low energy QCD~\cite{eta:1}. Precise measurements of their branching fractions (BFs) are important for a wide variety of physics topics. For example, the decay widths of $\eta/\eta'\to\gamma\gamma$ are related to the quark content of the two mesons~\cite{eta:2}, the BFs of $\eta/\eta'\to 3\pi$ decays can provide valuable information on light quark masses~\cite{eta:3}, the BFs of $\eta/\eta'\to\pi^+\pi^-\gamma$ decays are related to details of chiral dynamics~\cite{eta:4}, and the BFs of some rare decays of the $\eta$ and $\eta'$ can test fundamental QCD symmetries~\cite{eta:5} and probe for physics beyond the standard model~\cite{eta:6}. As the BFs of the rare decays are obtained via normalization to the dominant decay modes, a precise determination of the BFs of the dominant decay modes of the $\eta$ and $\eta'$ is essential. 

In Ref.~\cite{eta:0,etap:0}, we developed an approach to measure the absolute BFs of the exclusive decays of the $\eta/\eta^\prime$ meson using the data sample collected with the BESIII detector.
Taking advantage of the excellent momentum resolution of charged tracks in the MDC, photon conversions to $e^+e^-$ pairs provide a unique tool to reconstruct the inclusive photon spectrum from radiative $J/\psi$ decays. Take $J/\psi\rightarrow\gamma\eta(\eta^\prime)$ for example, Monte Carlo~(MC) study indicates that the energy resolution of the radiative photon could be improved by a factor of three using the photon conversion events. This enables us to tag the $\eta/\eta^\prime$ inclusive decays and then to measure the absolute BF of $J/\psi\rightarrow\gamma\eta(\eta^\prime)$, using
\begin{equation}
	Br(J/\psi \to \gamma \eta/\eta') = \frac{{{N^{\rm obs}_{J/\psi\rightarrow\gamma\eta/\eta^\prime}}}}{{N_{J/\psi}\cdot \varepsilon\cdot f}},
	\label{eq:gametapbf}
\end{equation}	
where $N^{\rm obs}_{J/\psi\rightarrow\gamma\eta/\eta^\prime}$ is the observed $\eta/\eta'$ yield, $\varepsilon$ is the detection efficiency obtained from MC simulation, and $N_{J/\psi }$ is the  number of $J/\psi$ events. The photon conversion process is simulated with GEANT4, and $f$ is a correction factor to account for the difference in the photon conversion efficiencies between data and MC simulation.

After the $\eta/\eta'$ inclusive measurement, we present precision measurements of $\eta$ decays to $\pi^0\pi^0\pi^0$, $\pi^+\pi^-\pi^0$, $\pi^+\pi^-\gamma$, $\gamma\gamma$, and $\eta^\prime$ decays to $\gamma\pi^+\pi^-$, $\eta\pi^+\pi^-$, $\eta\pi^0\pi^0$, $\gamma\omega$,  $\gamma\gamma$, again using $J/\psi$ decays to $\gamma \eta/\eta'$, but with the radiative photon directly detected by the EMC to improve the statistics. With the help of Eq.~(\ref{eq:gametapbf}), the BF for each $\eta^\prime$ exclusive decay is then calculated using
\begin{equation}
	Br(\eta/\eta' \to X) = \frac{{{N_{\eta/\eta' \to X }^{\rm obs}}}}{{{\varepsilon _{\eta/\eta' \to X }}}} \cdot \frac{{{\varepsilon}}}{{{N^{\rm obs}_{J/\psi\rightarrow\gamma\eta/\eta^\prime}}}} \cdot f,
	\label{eq:gamgambf}
\end{equation}
where $N_{\eta/\eta' \to X}^{\rm obs}$ is the number of signal events obtained from a fit to data and $\varepsilon_{\eta/\eta' \to X }$ is the MC-determined reconstruction efficiency. 

The measured BF of $J/\psi\to\gamma\eta(\eta')$ is $(1.067\pm0.005\pm0.023)\times 10^{-3}$ ($(5.27\pm0.03\pm0.05)\times 10^{-3}$), which is in agreement with the world average value, $(1.085\pm 0.018)\times 10^{-3}$ $((5.25\pm 0.07)\times 10^{-3}$)~\cite{PDG}, but with a significantly improved precision. In addition, we also give the relative BFs for $\eta$ and $\eta'$ decays as presented in
Tab.~\ref{tab:etasum} and Tab.~\ref{tab:etapsum} respectively.

\begin{table*}[htbp]
 \begin{center}
 \caption{Summary of BFs for $\eta$ decays and the comparison with previous results. The first error is statistical and the second systematic.}
 \label{tab:etasum}
	{\begin{tabular}{lccc}\hline\hline
	&  \multicolumn{3}{c}{$Br(\eta\to X)~(\%)$}\\\cline{2-4}
	$X$ & This Work & CLEO & PDG\\\hline
	$\gamma\gamma$ & 39.86$\pm$ 0.04$\pm$0.99 & 38.45$\pm$0.40$\pm$0.36 & 39.41$\pm$0.20\\
	$\pi^0\pi^0\pi^0$ & 31.96$\pm$0.07$\pm$0.84 & 34.03$\pm$0.56$\pm$0.49 & 32.68$\pm$0.23\\
	$\pi^+\pi^-\pi^0$ & 23.04$\pm$0.03$\pm$0.54 & 22.60$\pm$0.35$\pm$0.29 & 22.92$\pm$0.28\\
	$\pi^+\pi^-\gamma$ & 4.38$\pm$0.02$\pm$0.10& 3.96$\pm$0.14$\pm$0.14 & 4.22$\pm$0.08\\
	\hline\hline
		\end{tabular}}
	\end{center}
\end{table*}

\begin{table*}[htbp]
\begin{center}
\caption{Summary of the measured BFs for $\eta'$ decays. $Br$ is the determined BF.}
	\label{tab:etapsum}
	\begin{tabular}{lcccc}\hline\hline
    &  \multicolumn{2}{c}{$Br(\eta\to X)~(\%)$} &\multicolumn{2}{c}{$Br/Br(\eta' \to \eta \pi^{+} \pi^{-})$}\\\cline{2-5}
	Decay Mode&This measurement &PDG &This measurement	&CLEO \\ \hline
	$\eta'\to\gamma\pi^{+}\pi^{-}$
	&29.90$\pm$0.03$\pm$0.55 &28.9$\pm$0.5 &0.725$\pm$0.002$\pm$0.010 &0.677$\pm$0.024$\pm$0.011\\
	$\eta'\to\eta\pi^{+}\pi^{-}$			&41.24$\pm$0.08$\pm$1.24 &42.6$\pm$0.7 &...  & ...\\
	$\eta'\to\eta\pi^{0}\pi^{0}$	&21.36$\pm$0.10$\pm$0.92 &22.8$\pm$0.8 &0.518$\pm$0.003$\pm$0.021 &0.555$\pm$0.043$\pm$0.013\\
	$\eta'\to\gamma\omega$				&2.489$\pm$0.018$\pm$0.074	&2.62$\pm$0.13 &0.0604$\pm$0.0005$\pm$0.0012			&0.055$\pm$0.007$\pm$0.001\\
	$\eta'\to\gamma\gamma$          	&2.331$\pm$0.012$\pm$0.035 &2.22$\pm$0.08 	&0.0565$\pm$0.0003$\pm$0.0015			&0.053$\pm$0.004$\pm$0.001\\
	\hline \hline
	\end{tabular}
	\end{center}
\end{table*}

\section{Conclusion}
The BESIII collaboration has produced fruitful results related with light meson decays, including the studies of the decay dynamics, tests of discrete symmetries, searches for rare decays, and many other interesting results not covered in this proceeding. The BESIII experiment has
accumulated 10 billion $J/\psi$ events in total, which is a unique world wide sample, allows to study the light mesons with unprecedented statistics. Ongoing analyses will produce more precise results in the next years.

\end{document}